\documentclass[a4paper,11pt]{article}
\pdfoutput=1 % if your are submitting a pdflatex (section if you have
\usepackage{jinstpub} % for details on the use of the package, please
\usepackage{graphicx}
\usepackage{caption}
\usepackage{subcaption}
\usepackage{amsmath}   
\usepackage{xspace}

\usepackage[printwatermark]{xwatermark}
\usepackage{xcolor}
\usepackage{lipsum}
\usepackage{tikz}
\usepackage{lineno}
\usepackage{lipsum}
\usepackage{soul}

\title{Preliminary results from the cosmic data taking of the BESIII cylindrical GEM detectors}
\author{R. Farinelli}
\author{on behalf of the BESIII CGEM-IT working group}
\affiliation{INFN, Sezione di Ferrara, via G. Saragat 1, 44122 Ferrara, Italy}
\emailAdd{rfarinelli@fe.infn.it}

\abstract{BESIII is a multipurpose spectrometer optimized for physics in the tau-charm energy region. Both detector and accelerator are undergoing an upgrade program, that will allow BESIII to run until 2029. A major upgrade is the replacement of the inner drift chamber with a new detector based on Cylindrical Gas Electron Multipliers to improve both the secondary vertex reconstruction and the radiation tolerance. The CGEM-IT will be composed of three coaxial layers of cylindrical triple GEMs, operating in an Ar$\,$+$\,$iC$_4$H$_{10}$ (90:10) gas mixture with field and gain optimized to maximize the spatial resolution. The new detector is readout with innovative TIGER electronics produced in 110$\,$nm CMOS technology. The front end is a custom designed 64$\,$channel ASIC featuring a fully digital output and operated in trigger-less mode. It can provide analog charge and time measurements with a TDC time resolution better than 100$\,$ps, that will allow to operate in $\upmu$TPC mode. With planar prototypes, we measured an unprecedented spatial resolution below 150$\,\upmu$m in a 1$\,$Tesla magnetic field in a wide range of incident angles of the incoming particle. Before the installation inside BESIII, foreseen in 2021, a long standalone data taking is ongoing at the Institute of High Energy Physics in Beijing; currently, the first two cylindrical chambers are available for the test, and are used to complete the integration between the detector and the electronics and to assess the required performance. In this proceeding a description of the CGEM-IT project, the TIGER features and performance, and the results of the analysis of first cosmic ray data taking will be presented. Focus will be given on the strip analysis, from which it is possible to measure the basic properties of the detector, and the cluster analysis, where a comparison with the results with planar prototypes will be discussed. The first preliminary results on efficiency and spatial resolution will be also presented.}

\arxivnumber{XXX}
\proceeding{INTERNATIONAL CONFERENCE INSTRUMENTATION FOR COLLIDING BEAM PHYSICS
24-28 FEBRUARY 2020\\
BUDKER INSTITUTE OF NUCLEAR PHYSICS, NOVOSIBIRSK, RUSSIA}

\begin{document}
\maketitle
\flushbottom
\section{Introduction}
The BEijing Spectrometer (BESIII,$\,$\cite{BESIII}) is an apparatus composed of several sub-detectors that measures the properties of the particles in the energy range from 2 to 4.6$\,$GeV. The lepton beams are generated by the Beijing Electron Positron Collider II (BEPCII) and their collision is operated at a luminosity of 10$^{33}\,$cm$^{-2}$s$^{-1}$. At present, the momentum of charged particles is measured by the Main Drift Chamber (MDC)  built around the beryllium beam pipe. Outside the MDC, the time-of-flight system identifies the particle type and the Electro-Magnetic Calorimeter their energy. The outermost detector is the MUon Counter built in the yoke of the 1$\,$T superconducting magnet. In 2019, an upgrade of the BEPCII machine and some sub-detectors of BESIII has started. The beam energy is boosted to explore the region above the $\Lambda_c$ threshold and a top-up injection mode is used to increase the luminosity. From the BESIII side, the TOF is upgraded with a new technology of Multi Resistive Plate Chamber and the Inner Drift Chamber needs a replacement due to aging effect. The Cylindrical Gas Electron Multiplier (CGEM) is a suitable technology to upgrade the BESIII Inner Tracker (IT).

\vspace{0.5cm}
\section{The Cylindrical GEM Inner Tracker}
\begin{figure}[tbp]
    \centering
        \includegraphics[width=0.4\textwidth]{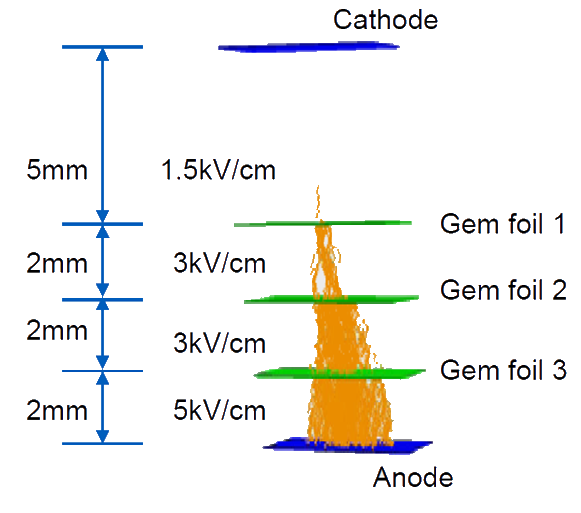} 
        \,\,\,\,\,\,\,\,\,\,\,\,\,\,\,\,\,\,\,\,\,\,\,\,    
        \includegraphics[width=0.35\textwidth]{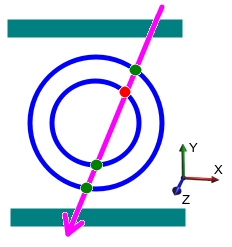}
        \caption{On the left a representation of a triple GEM detector. The blue foils are cathode and anode, while the green ones are the three amplification stages. The orange lines represent the path of the electrons. The applied electric field and the distance between each electrodes are shown. On the right a scheme of the setup with the two CGEM layers (blue circles) and the scintillating bars (light blue boxes). The pink line is the cosmic ray, the four points are the hits on the CGEMs: the three green dots are used for the track reconstruction and the red for the characterization of the detector.}
        \label{fig:triple}
\end{figure}

The CGEM-IT composed up by three layers of Cylindrical GEM. Each one provides a three dimensional reconstruction of the particle position and thanks to its shape it can cover the 93$\%$ solid angle around the beam pipe. The GEM technology$\,$\cite{GEM} is a Micro Pattern Gas Detector widely used in high energy physics to detect the primary ionization of the interacting charged particles. The amplification of the signal is given by the GEM foils: a 50$\,\upmu$m thick kapton foil with copper faces and bi-conical holes. An intense electric field between the GEM faces amplifies the number of electrons entering the holes. Several stages of amplification guarantee an high gain of the detector with a low discharge probability$\,$\cite{discharge}. The full geometry of a single cylindrical layer is given by a cathode, three GEM foils and the readout plane, as shown in Fig.$\,$\ref{fig:triple}, left. Two electrodes are separated by a gap of 2$\,$mm but the cathode by 5$\,$mm. The mechanical structure is very light and the full interaction length of each layer is below 0.5$\,\%$ of X$_0$.

\subsection{Signal readout}
\begin{figure}[tbp]
    \centering
        \includegraphics[width=0.8\textwidth]{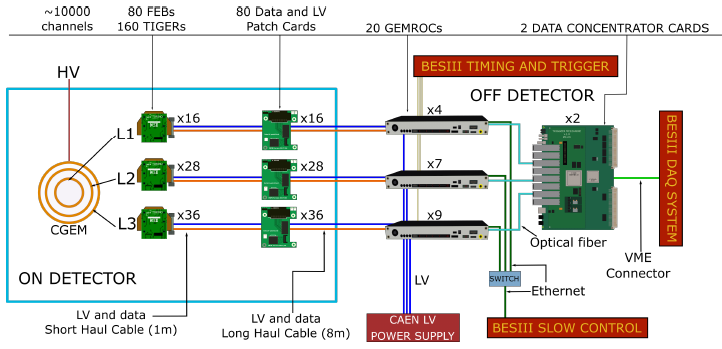} \caption{Readout chain from the CGEM-IT to the BESIII DAQ system. The {\it on detector} electronics are the chips connected to the CGEM anode plane and the patch cards. The {\it off detector} electronics manages the input/output of the TIGER through the GEMROC and the data concentrator and it is connected the the power supply, the slow control and the DAQ.}
        \label{fig:chain}
\end{figure}

The readout plane is segmented by longitudinal ({\it X}) and stereo ({\it V}) strips. The pitch is 660$\,\upmu$m and the angle between the strips varies layer by layer. The length of the strips is fixed for the longitudinal ones, while the stereo varies from few millimeter up a tens of centimeters. A signal of about 100$\,$fC per m.i.p. is collected on the strips and the TIGER (Torino Integrated Gem Electronics for Readout) electronics$\,$\cite{TIGER} is used to measure the time and the collected charge. Two ASICs are installed on a Front-End Board (FEB) to read the signal from 128$\,$strips. The chip has two separate branches to extract time and charge information. It is possible to operate with thresholds on both branches to optimize the readout performance. The thresholds depend on the strips capacitance. The charge can be measured with two methods: Sample and Hold (S/H) and Time-Over-Threshold (ToT). The S/H provides a linear measurement of the signal amplitude in a dynamic range from 1 to 40-50$\,$fC per channel; the ToT converts the time spent by the signal over the threshold into charge information. The chip has been characterized: calibration curves with external test pulse are defined for the S/H and ToT modes, the baselines of each channel is equalized. The time-walk effects have been evaluated as a function of the signal amplitude and the channel threshold. Every signal crossing the threshold on both branches is digitized and transmitted to the off detector electronics. Once the trigger is sent to the readout chain, the data measured by each chip are collected. The GEM Read Out Card (GEMROC) manages the low voltage of the FEB, the chip configuration and the data collection. The final design is summarized in Fig.$\,$\ref{fig:chain}, it will have 80$\,$FEB, 20$\,$GEMROC and two data concentrator cards to send the whole CGEM-IT outputs to the BESIII DAQ system. 

\subsection{CGEMBOSS}
The reconstruction and the analysis of the data is implemented in the BESIII Offline Software System (BOSS,$\,$\cite{BOSS}). This environment takes care of the geometry description of the detector, the simulation and the reconstruction of real data. Within the BOSS environment, the measurements from the CGEM-IT are used in the global track finder algorithms and in the track fitting procedure to measure the particle path. The results shown in this proceeding use a preliminary version of the calibration and alignment. 

\subsection{Signal reconstruction}
Contiguous firing strips are clusterized and their charge and time information are used to characterize the signal. The digitized signal of each strip is labelled as {\it hit}. The charge of all the strips of the cluster is proportional to the deposited energy. The position is reconstructed with two methods: the Charge Centroid (CC) and the micro-Time Projection Chamber ($\upmu$TPC,$\,$\cite{triplegem}). The first averages the position of each strip of the cluster by weighting it by its charge while the second associates to each strip a bi-dimensional point and it uses a linear fit on the points to extrapolate the particle position. The clusterization is performed on longitudinal and stereo strips. The combination of the two gives a bi-dimensional reconstruction.

\begin{figure}[tbp]
    \centering
        \includegraphics[width=0.45\textwidth]{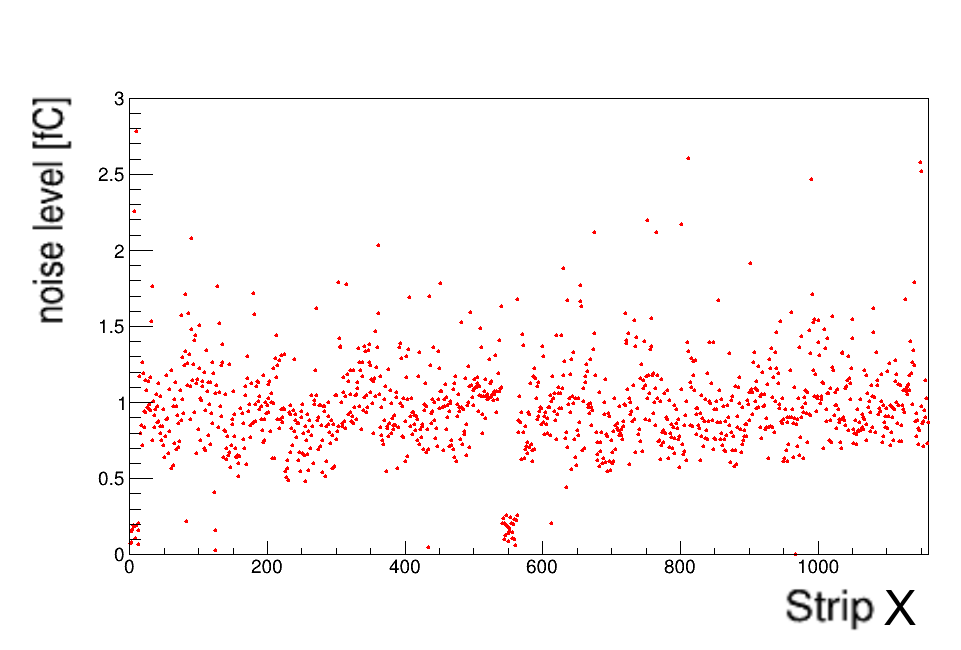} 
        \,\,\,\,\,\,\,\,\,\,\,\,\,\,\,\,
        \includegraphics[width=0.45\textwidth]{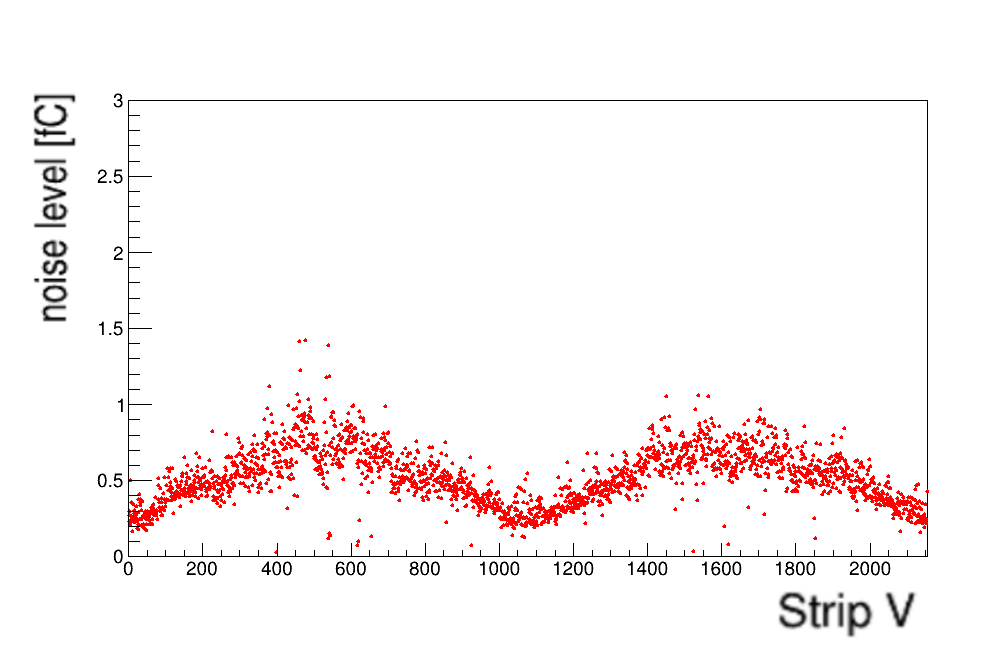}
        \caption{Noise level of the strips for the outer-most layer. On the left the longitudinal strips are reported and on the right the stereo ones. The noise depends mainly on the strip length.}
        \label{fig:noise}
\end{figure}

\vspace{0.5cm}
\section{Settings and setup configurations}
Two of the three CGEM layers have been installed together in Beijing, at the Institute of High Energy Physics (IHEP) experimental test area. The cylinders have an inner radius of 76.9$\,$mm and 121.4$\,$mm, while their length is 532$\,$mm and 690$\,$mm in the active area. The pitch size of the strips is 660$\,\upmu$m, therefore the total number of channel instrumented at present is about five thousands. The gas mixture used inside the CGEM is Ar$\,$+$\,$iC$_4$H$_{10}$ (90:10) to provide about 55 electrons from ionization per m.i.p. in 5$\,$mm. The total high voltage on the three GEM electrodes is 830$\,$V that corresponds to 12'000 detector gain, while the electric fields between the electrodes are set to 1.5/3/3/5$\,$kV/cm\footnote{electric fields in the gaps between the electrodes from the cathode to the anode.}.
Once the full readout chain is connected to the detectors, a threshold scan is performed to measure the noise level on each channel. On-chip test pulse is used to measure the width of the noise distribution and a different level of noise as a function of the strip length is measured. In detail, as shown in Fig.$\,$\ref{fig:noise}, the longitudinal strips with same length have a noise of about 1$\,$fC but in stereo strips it varies from 0.2 to 1$\,$fC. The threshold level applied to the channel is set in order to have the same noise rate on all the channels around 8$\,$kHz.
Cosmic interacting with both CGEM layer are selected by the GEMROC modules trigger-matching algorithm and used in the study of the performance.
\vspace{0.5cm}
\section{Analysis method}
A cosmic ray data taking is ongoing at IHEP to integrate the detector with the electronics, to develop the software algorithms needed in the reconstruction of the events and to characterize the performance of the IT. The cosmic ray interacting with the setup generates a signal on both halves of the cylinders: this generates at least four points, two per CGEM layer. The readout planes are divided in two halves, top and bottom. Three points are used to track while the fourth is tested. A sketch of the setup is shown in Fig.$\,$\ref{fig:triple}. right. The residual distribution of the difference between the expected position from the trackers and the measured one from the CGEM part under test is measured. Only Charge Centroid (CC) is used for the following results. This study is performed for all the four parts of the CGEM-IT with a permutation of test and trackers planes. The residual distribution strictly depends on the trackers, on the pattern recognition algorithms and of course on the detector itself. At present, a cut on the trackers fit $\chi^2$ permits to distinguish the real signal events from the noise event wrongly selected. This technique reduces the statistics of the data but it allows easily to reject events where the cluster selection inefficiency impacts on the measurement of the performance, {\it i.e.} combinatorial noise. Once the $\chi^2$ cut is selected, a rough alignment is performed with a shift in the {\it Z} coordinate along the cylinder axis and a rotation around the cylinder axis. This allows to measure a residual distribution of the {\it XY} and {\it Z} coordinates centered in zero. 

\begin{figure}[tbp]
    \centering
        \includegraphics[width=0.32\textwidth]{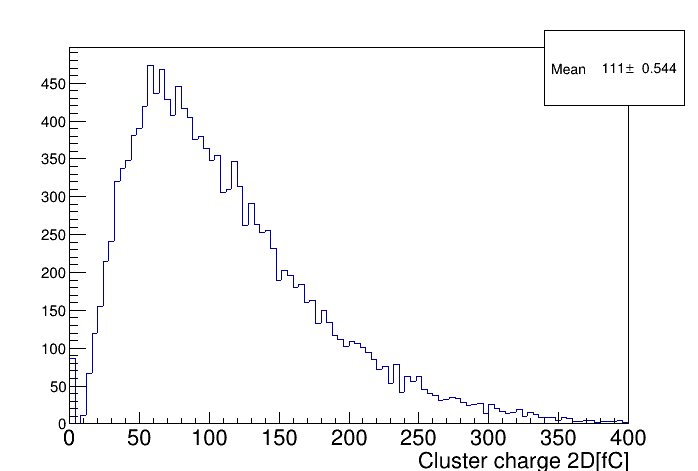} 
        \includegraphics[width=0.32\textwidth]{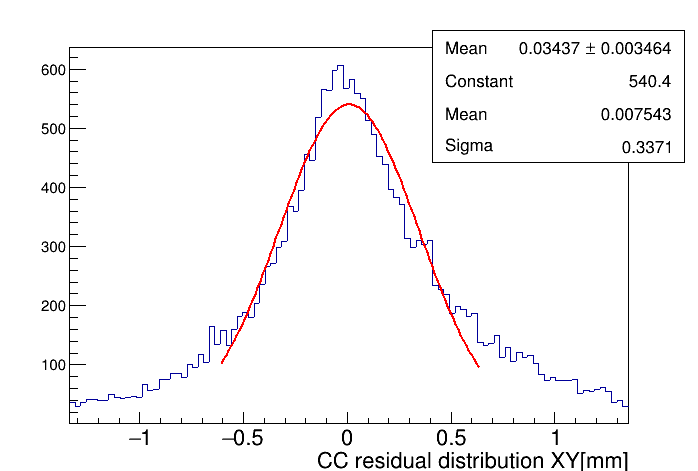} 
        \includegraphics[width=0.32\textwidth]{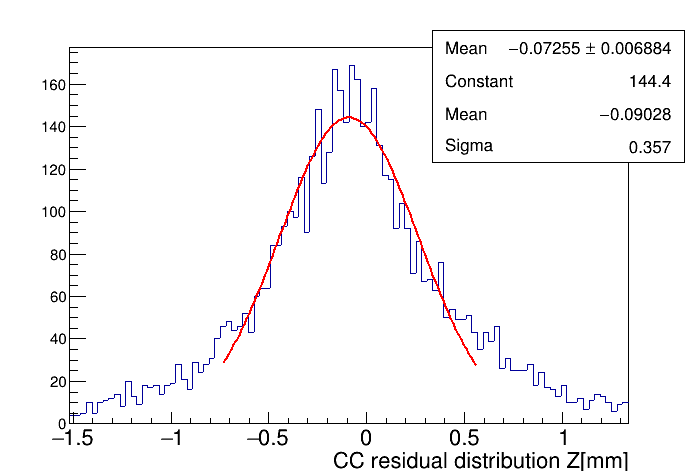} 
        \caption{On the left the charge distribution of the bi-dimensional clusters is shown. A mean value of 110$\,$fC is reported. On the center and right the residual distribution on the {\it XY} and {\it Z} planes after the alignment procedures and the $\chi^2$ cut are shown. A Gaussian fit is performed on the core distribution and a sigma of about 350$\,\upmu$m is reported.}
        \label{fig:res1}
\end{figure}

\vspace{0.5cm}
\section{Performance measurement}
\begin{figure}[tbp]
    \centering
        \includegraphics[width=0.45\textwidth]{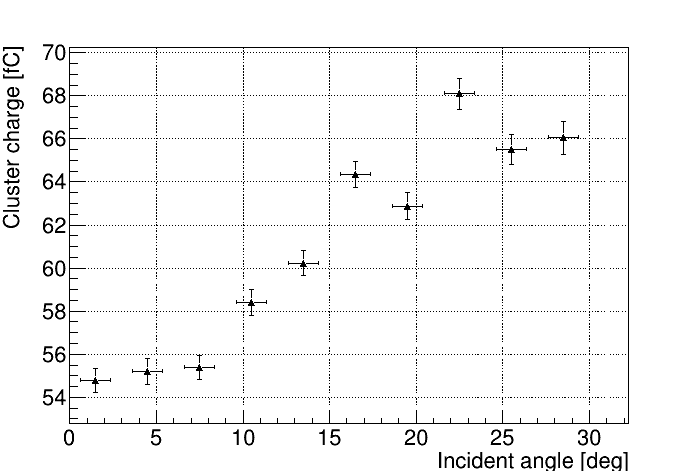} 
        \includegraphics[width=0.45\textwidth]{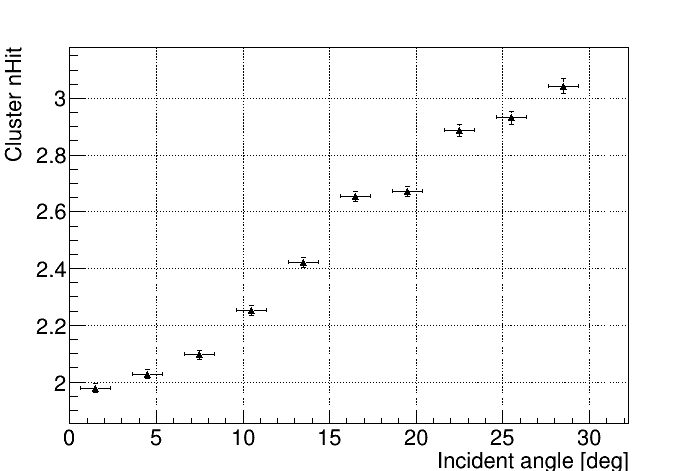}\\ 
        \includegraphics[width=0.45\textwidth]{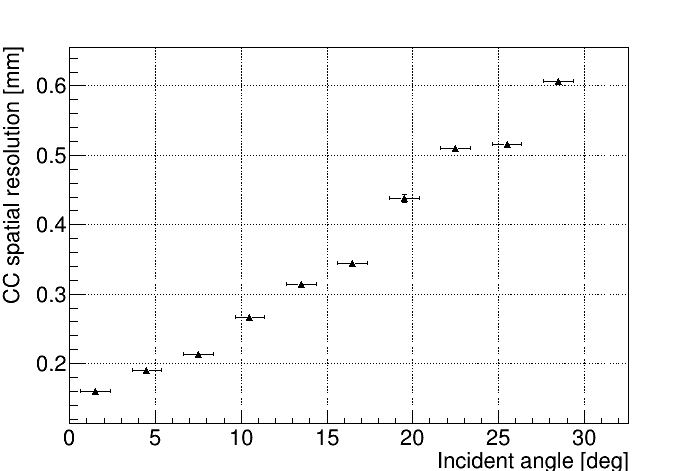} 
        \includegraphics[width=0.45\textwidth]{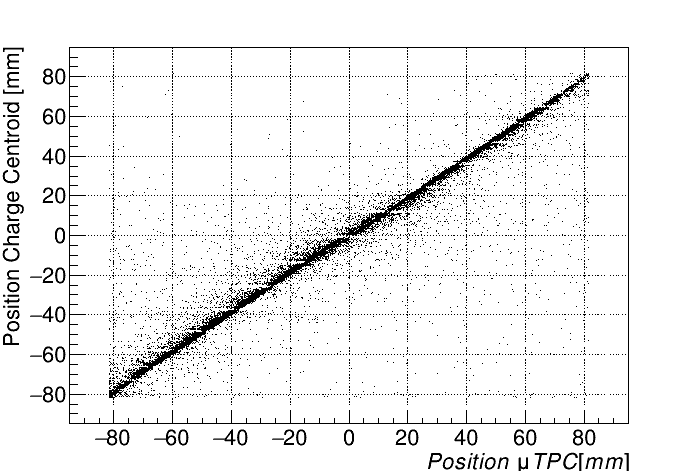}
        \caption{On the top the cluster charge (left) and the cluster size (right) of the {\it X} cluster as a function of the incident angle is shown. On the bottom left the charge centroid spatial resolution as a function of the incident angle. At 0$^\circ$ the best resolution is achieved. The incident angle is the angle between the particle track and the surface normal vector in the {\it XY} plane. On the bottom right is reported the correlation between the CC and the $\upmu$TPC measurements is reported.}
        \label{fig:res2}
\end{figure}

The analysis is performed for each one of the four parts. The clusters of strips within five sigma in the residual distribution are identified as signal then cluster charge, size, efficiency and resolution are evaluated. An average charge of 110$\,$fC has been measured for a bi-dimensional cluster, in agreement with the HV setting, as shown in the left part of Fig.$\,$\ref{fig:res1}. The efficiency is evaluated as the ratio between the number of clusters within five sigma of the residual distribution and the number of good tracks. A value around 90$\,\%$ is measured for bi-dimensional clusters. This result is affected by the present status of the setup and the reconstruction: the readout chain and the pattern recognition bias this results to a lower value. The sigma of the residual distribution is about 350$\,\upmu$m both on the {\it XY} plane than in {\it Z} direction, as reported in Fig.$\,$\ref{fig:res1} center and right. To understand this value it is important to study the CC as a function of the incident angle. The CC is an algorithm with good performance if the charge distribution is Gaussian, but for a large incident angle this is no more possible and it degrades. As expected, the cluster charge and the cluster size  increase with the angle between the track and the normal to the cylinder surface. The results from the longitudinal strips are reported: for events with a small cluster size, the sigma of the CC distribution is below 200$\,\upmu$m, as shown in Fig.$\,$\ref{fig:res2}, as a function of the incident angle. When the track has a large incident angle, the CC is no more efficient and the $\upmu$TPC algorithm has to be used. The $\upmu$TPC is a more delicate algorithms and it needs dedicated calibration studies. Up to now a preliminary reconstruction with the $\upmu$TPC has been performed and the results are in agreement with the CC. See Fig.$\,$\ref{fig:res2} bottom right.

\vspace{0.5cm}
\section{Conclusions}
The integration between two layers of cylindrical GEM and a full readout chain based on TIGER electronics is under study. The optimization of the setup and the software is ongoing. Preliminary results of the CGEM-IT status are analyzed: the noise level on the channels is flat for longitudinal strips, while on the stereo ones it follows the strip length. The global efficiency is around 90$\%$ and this reflects the present status of the hardware and the software tools. Improvements on both sides will be implemented to optimize the performance of the detector. The cluster charge and size are good and their dependency on the incident angle is reasonable. The spatial resolution up to now has been evaluated with the CC only and the results copy the planar triple-GEM behavior$\,$\cite{triplegem}: for orthogonal tracks the resolution is below 200$\,\upmu$m in the {\it XY} plane, while in the {\it Z} direction it is three times better than the current BESIII-MDC resolution. It is needed to remove the contribution of the tracking system and the multiple scattering of the mechanical structure to measure the spatial resolution from the shown results. The reconstruction of the first $\upmu$TPC event with a CGEM and a TIGER chip has been shown and an optimization of this algorithm will be implemented with further upgrades of the calibrations.

\vspace{0.5cm}
\acknowledgments
This work is supported by the Italian Institute of Nuclear Physics (INFN). \\
The research leading to these results has been performed within the FEST Project, funded by the European Commission in the call RISE-MSCA-H2020-2020.


\end{document}